\documentclass[aps,twocolumn,superscriptaddress,reprint]{revtex4-2}
\usepackage[T1]{fontenc}
\usepackage[colorinlistoftodos]{todonotes}
\usepackage{newtxtext,newtxmath,booktabs,siunitx} 
\usepackage{dcolumn}
\usepackage{bm}
\usepackage{float}
\usepackage{subcaption}
\usepackage[]{graphicx}  
\usepackage{booktabs}
\usepackage{tabularx}
\usepackage{amsmath}
\usepackage{xcolor}
\usepackage{ulem}
\usepackage[colorlinks,citecolor=blue,urlcolor=blue,linkcolor=blue]{hyperref}

\DeclareSIUnit\eVperc{\eV\per\clight}
\DeclareSIUnit\clight{\text{\ensuremath{c}}}
\DeclareSIUnit\MeVA{A \mega\eV }

\newcommand{\spirit}{S$\pi$RIT }

\newcolumntype{C}{>{\centering\arraybackslash}X}

\newcommand{\comment}[1]{}
\newcommand{\Rf}{$R_{c}$~}

\makeatletter
\renewcommand\@makecaption[2]{%
  \par
  \vskip\abovecaptionskip
  \begingroup
   \small\rmfamily
    \begingroup
     \samepage
     \flushing
     \let\footnote\@footnotemark@gobble
     \@make@capt@title{#1}{#2}\par
    \endgroup
  \endgroup
  \vskip\belowcaptionskip
}
\makeatother

\graphicspath{ {images/} }

\usepackage[mathlines]{lineno}

\begin{document}





\title{ Large amplification of the isospin-dependence of proton emitting source size in  { nuclear collisions induced by radioactive ion beams} : A signal of n-p correlation }





\author{Y. J.~Wang}
\altaffiliation{Corresponding author}
\email{wyj25@buaa.edu.cn}

\thanks{Present address: School of Physics, Beihang University, Beijing 100191, China.}
\affiliation{Department of Physics, Tsinghua University, Beijing 100084, China}
\affiliation{RIKEN Nishina Center, Hirosawa 2-1, Wako, Saitama 351-0198, Japan}

\author{C. K.~Tam}
\affiliation{Department of Physics, Western Michigan University,
Kalamazoo, Michigan 49008, USA}

\author{Z.~G.~Xiao}
\altaffiliation{Corresponding author}
\email{xiaozg@tsinghua.edu.cn}
\affiliation{Department of Physics, Tsinghua University, Beijing 100084, China}
\affiliation{Center for High Energy Physics, Tsinghua University, Beijing 100084, China}

\author{W.G.~Lynch}
\altaffiliation{Corresponding author}
\email{lynch@nscl.msu.edu}
\affiliation{National Superconducting Cyclotron Laboratory, Michigan State University, East Lansing, Michigan 48824, USA}
\affiliation{Department of Physics, Michigan State University, East Lansing, Michigan 48824, USA}

\author{C.Y.~Tsang}
\affiliation{National Superconducting Cyclotron Laboratory, Michigan State University, East Lansing, Michigan 48824, USA}
\affiliation{Department of Physics, Michigan State University, East Lansing, Michigan 48824, USA}

\author{J.~Barney}
\affiliation{National Superconducting Cyclotron Laboratory, Michigan State University, East Lansing, Michigan 48824, USA}
\affiliation{Department of Physics, Michigan State University, East Lansing, Michigan 48824, USA}

\author{G.~Jhang}
\affiliation{National Superconducting Cyclotron Laboratory, Michigan State University, East Lansing, Michigan 48824, USA}

\author{J.~Estee}
\affiliation{National Superconducting Cyclotron Laboratory, Michigan State University, East Lansing, Michigan 48824, USA}
\affiliation{Department of Physics, Michigan State University, East Lansing, Michigan 48824, USA}

\author{M.B.~Tsang}
\altaffiliation{Corresponding author}
\email{tsang@nscl.msu.edu}
\affiliation{National Superconducting Cyclotron Laboratory, Michigan State University, East Lansing, Michigan 48824, USA}
\affiliation{Department of Physics, Michigan State University, East Lansing, Michigan 48824, USA}

\author{R.~S.~Wang}
\affiliation{School of Radiation Medicine and Protection, Soochow University, Suzhou, China}

\author{M.~Kaneko}
\affiliation{RIKEN Nishina Center, Hirosawa 2-1, Wako, Saitama 351-0198, Japan}
\affiliation{Department of Physics, Kyoto University, Kita-shirakawa, Kyoto 606-8502, Japan}

\author{J.W.~Lee}
\affiliation{Department of Physics, Korea University, Seoul 02841, Republic of Korea}

\author{J. ~Park}
\affiliation{Department of Physics, Korea University, Seoul 02841, Republic of Korea}

\author{Z. ~Chajęcki}
\altaffiliation{Corresponding author}
\email{zbigniew.chajecki@wmich.edu}
\affiliation{Department of Physics, Western Michigan University,
Kalamazoo, Michigan 49008, USA}

\author{G. ~Verde}
\altaffiliation{Corresponding author}
\email{giuseppe.verde@ct.infn.it}
\affiliation{INFN - Sezione di Catania, 95123, Catania, Italy}

\author{T.~Isobe}
\altaffiliation{Corresponding author}
\email{isobe@riken.jp}
\affiliation{RIKEN Nishina Center, Hirosawa 2-1, Wako, Saitama 351-0198, Japan}

\author{M.~Kurata-Nishimura}
\affiliation{RIKEN Nishina Center, Hirosawa 2-1, Wako, Saitama 351-0198, Japan}

\author{T.~Murakami}
\affiliation{RIKEN Nishina Center, Hirosawa 2-1, Wako, Saitama 351-0198, Japan}
\affiliation{Department of Physics, Kyoto University, Kita-shirakawa, Kyoto 606-8502, Japan}

\author{D.S.~Ahn}
\affiliation{RIKEN Nishina Center, Hirosawa 2-1, Wako, Saitama 351-0198, Japan}

\author{L.~Atar}
\affiliation{Institut f\"ur Kernphysik, Technische Universit\"at Darmstadt, D-64289 Darmstadt, Germany}
\affiliation{GSI Helmholtzzentrum f\"ur Schwerionenforschung, Planckstrasse 1, 64291 Darmstadt, Germany}

\author{T.~Aumann}
\affiliation{Institut f\"ur Kernphysik, Technische Universit\"at Darmstadt, D-64289 Darmstadt, Germany}
\affiliation{GSI Helmholtzzentrum f\"ur Schwerionenforschung, Planckstrasse 1, 64291 Darmstadt, Germany}

\author{H.~Baba}
\affiliation{RIKEN Nishina Center, Hirosawa 2-1, Wako, Saitama 351-0198, Japan}

\author{K.~Boretzky}
\affiliation{GSI Helmholtzzentrum f\"ur Schwerionenforschung, Planckstrasse 1, 64291 Darmstadt, Germany}

\author{J.~Brzychczyk}
\affiliation{Faculty of Physics, Astronomy and Applied Computer Science, Jagiellonian University, Krak\'ow, Poland}

\author{G.~Cerizza}
\affiliation{National Superconducting Cyclotron Laboratory, Michigan State University, East Lansing, Michigan 48824, USA}

\author{N.~Chiga}
\affiliation{RIKEN Nishina Center, Hirosawa 2-1, Wako, Saitama 351-0198, Japan}

\author{N.~Fukuda}
\affiliation{RIKEN Nishina Center, Hirosawa 2-1, Wako, Saitama 351-0198, Japan}

\author{I.~Gasparic}
\affiliation{Division of Experimental Physics, Rudjer Boskovic Institute, Zagreb, Croatia}
\affiliation{RIKEN Nishina Center, Hirosawa 2-1, Wako, Saitama 351-0198, Japan}
\affiliation{Institut f\"ur Kernphysik, Technische Universit\"at Darmstadt, D-64289 Darmstadt, Germany}

\author{B.~Hong}
\affiliation{Department of Physics, Korea University, Seoul 02841, Republic of Korea}

\author{A.~Horvat}
\affiliation{Institut f\"ur Kernphysik, Technische Universit\"at Darmstadt, D-64289 Darmstadt, Germany}
\affiliation{GSI Helmholtzzentrum f\"ur Schwerionenforschung, Planckstrasse 1, 64291 Darmstadt, Germany}

\author{K.~Ieki}
\affiliation{Department of Physics, Rikkyo University, Nishi-Ikebukuro 3-34-1, Tokyo 171-8501, Japan}

\author{N.~Inabe}
\affiliation{RIKEN Nishina Center, Hirosawa 2-1, Wako, Saitama 351-0198, Japan}

\author{Y.J.~Kim}
\affiliation{Rare Isotope Science Project, Institute for Basic Science, Daejeon 34047, Republic of Korea}

\author{T.~Kobayashi}
\affiliation{Department of Physics, Tohoku University, Sendai 980-8578, Japan}

\author{Y.~Kondo}
\affiliation{Department of Physics, Tokyo Institute of Technology, Tokyo 152-8551, Japan}

\author{P.~Lasko}
\affiliation{Faculty of Physics, Astronomy and Applied Computer Science, Jagiellonian University, Krak\'ow, Poland}

\author{H.~S.~Lee}
\affiliation{Rare Isotope Science Project, Institute for Basic Science, Daejeon 34047, Republic of Korea}

\author{Y.~Leifels}
\affiliation{GSI Helmholtzzentrum f\"ur Schwerionenforschung, Planckstrasse 1, 64291 Darmstadt, Germany}

\author{J.~\L{}ukasik}
\affiliation{Institute of Nuclear Physics PAN, ul. Radzikowskiego 152, 31-342 Krak\'ow, Poland}

\author{J.~Manfredi}
\affiliation{National Superconducting Cyclotron Laboratory, Michigan State University, East Lansing, Michigan 48824, USA}
\affiliation{Department of Physics, Michigan State University, East Lansing, Michigan 48824, USA}

\author{A.~B.~McIntosh}
\affiliation{Cyclotron Institute, Texas A\&M University, College Station, Texas 77843, USA}

\author{P.~Morfouace}
\affiliation{National Superconducting Cyclotron Laboratory, Michigan State University, East Lansing, Michigan 48824, USA}

\author{T.~Nakamura}
\affiliation{Department of Physics, Tokyo Institute of Technology, Tokyo 152-8551, Japan}

\author{N.~Nakatsuka}
\affiliation{RIKEN Nishina Center, Hirosawa 2-1, Wako, Saitama 351-0198, Japan}
\affiliation{Department of Physics, Kyoto University, Kita-shirakawa, Kyoto 606-8502, Japan}

\author{S.~Nishimura}
\affiliation{RIKEN Nishina Center, Hirosawa 2-1, Wako, Saitama 351-0198, Japan}

\author{H.~Otsu}
\affiliation{RIKEN Nishina Center, Hirosawa 2-1, Wako, Saitama 351-0198, Japan}

\author{P.~Paw\l{}owski}
\affiliation{Institute of Nuclear Physics PAN, ul. Radzikowskiego 152, 31-342 Krak\'ow, Poland}

\author{K.~Pelczar}
\affiliation{Faculty of Physics, Astronomy and Applied Computer Science, Jagiellonian University, Krak\'ow, Poland}

\author{D.~Rossi}
\affiliation{Institut f\"ur Kernphysik, Technische Universit\"at Darmstadt, D-64289 Darmstadt, Germany}

\author{H.~Sakurai}
\affiliation{RIKEN Nishina Center, Hirosawa 2-1, Wako, Saitama 351-0198, Japan}

\author{C.~Santamaria}
\affiliation{National Superconducting Cyclotron Laboratory, Michigan State University, East Lansing, Michigan 48824, USA}

\author{H.~Sato}
\affiliation{RIKEN Nishina Center, Hirosawa 2-1, Wako, Saitama 351-0198, Japan}

\author{H.~Scheit}
\affiliation{Institut f\"ur Kernphysik, Technische Universit\"at Darmstadt, D-64289 Darmstadt, Germany}

\author{R.~Shane}
\affiliation{National Superconducting Cyclotron Laboratory, Michigan State University, East Lansing, Michigan 48824, USA}

\author{Y.~Shimizu}
\affiliation{RIKEN Nishina Center, Hirosawa 2-1, Wako, Saitama 351-0198, Japan}

\author{H.~Simon}
\affiliation{GSI Helmholtzzentrum f\"ur Schwerionenforschung, Planckstrasse 1, 64291 Darmstadt, Germany}

\author{A.~Snoch}
\affiliation{Nikhef National Institute for Subatomic Physics, Amsterdam, Netherlands}

\author{A.~Sochocka}
\affiliation{Faculty of Physics, Astronomy and Applied Computer Science, Jagiellonian University, Krak\'ow, Poland}


\author{T.~Sumikama}
\affiliation{RIKEN Nishina Center, Hirosawa 2-1, Wako, Saitama 351-0198, Japan}

\author{H.~Suzuki}
\affiliation{RIKEN Nishina Center, Hirosawa 2-1, Wako, Saitama 351-0198, Japan}

\author{D.~Suzuki}
\affiliation{RIKEN Nishina Center, Hirosawa 2-1, Wako, Saitama 351-0198, Japan}

\author{H.~Takeda}
\affiliation{RIKEN Nishina Center, Hirosawa 2-1, Wako, Saitama 351-0198, Japan}

\author{S.~Tangwancharoen}
\affiliation{Department of Physics, Faculty of Science, King Mongkut's University of Technology Thonburi, Bangkok, 10140, Thailand}
\affiliation{Center of Excellence in Theoretical and Computational Science (TACS-CoE), Faculty of Science, King Mongkut's University of Technology Thonburi, Bangkok, 10140, Thailand}

\author{H.~Toernqvist}
\affiliation{Institut f\"ur Kernphysik, Technische Universit\"at Darmstadt, D-64289 Darmstadt, Germany}
\affiliation{GSI Helmholtzzentrum f\"ur Schwerionenforschung, Planckstrasse 1, 64291 Darmstadt, Germany}

\author{Y.~Togano}
\affiliation{Department of Physics, Rikkyo University, Nishi-Ikebukuro 3-34-1, Tokyo 171-8501, Japan}

\author{S.~J.~Yennello}
\affiliation{Cyclotron Institute, Texas A\&M University, College Station, Texas 77843, USA}
\affiliation{Department of Chemistry, Texas A\&M University, College Station, Texas 77843, USA}

\author{Y.~Zhang}
\affiliation{School of Physics and Technology, Nanjing Normal University, Nanjing 210023, China}

\collaboration{\spirit collaboration}




\date{\today}

\begin{abstract}

We report proton-proton correlation function measurements in central  $^{132}$Sn+$^{124}$Sn  and $^{108}$Sn+$^{112}$Sn collisions at 270~MeV/nucleon. The proton emitting source sizes are extracted for the systems by using femtoscopic imaging technique. The fast dynamic core radius for the neutron-rich system is found to be $2.22 \pm 0.13\ \text{(stat.)} \pm 0.07\ \text{(syst.)}$~fm, which is approximately 24\% larger than that for the neutron-deficient system, $1.74 \pm 0.08\ \text{(stat.)} \pm 0.05\ \text{(syst.)}$~fm. This difference is an order of magnitude larger than the $\sim$3\% difference in the ground-state charge radii of the projectile nuclei. Transport model simulations based on mean-field dynamics cannot reproduce this amplification. The observation reveals a beyond-mean-field mechanism associated to short-range neutron-proton correlations, which dynamically enhance the proton emitting source in the neutron-rich environment. Our results demonstrate that heavy-ion collisions induced by radioactive ion beam, combined with femtoscopic precision, provide a new hadronic probe of short-range correlation, and that careful treatment of the beyond-mean-field interactions are required in modeling such processes. 

\end{abstract}


\maketitle


The ground-state nuclear charge radius $\langle r_{\text{ch}}^2 \rangle$ is a fundamental observable that encodes the nature of nucleon-nucleon interactions and the spatial organization of nucleons within the nucleus. Across isotopic chains, charge radii increase with mass as $R\propto A^{1/3}$, a trend primarily attributed to the mean field linked to the equation of state  (EoS) of nuclear matter \cite{Angeli:2013epw, Gustafsson:2025xpd}. High-precision laser spectroscopy of tin isotopes \cite{Gustafsson:2025xpd} has shown that the smooth evolution between the neutron shell closures $N=50$ and $N=82$ is dominated by this mean-field growth, while beyond-mean-field correlations become appreciable only on the neutron-deficient end. Similar neutron-proton ($np$) correlation effects appear in other isotopic chains. For calcium, dispersive optical model studies of $p+^{40}$Ca and $p+^{48}$Ca scattering reveal that protons exhibit stronger beyond-mean-field correlations in the neutron-rich isotope \cite{Charity:2006zb}. Further toward the neutron-rich side, the rapid increase of charge radii beyond $^{48}$Ca is driven by extended valence neutron orbitals \cite{Enciu:2022vox}. All these results suggest that $np$ correlations, well established in electron-nucleus scattering experiments \cite{Hen:2014nza, CLAS:2018yvt, CLAS:2018xvc},  play a role in shaping nuclear charge radii \cite{MILLER2019360}.

An interesting but unresolved question is whether the $N/Z$-dependent variation of the charge radius persists under the extreme conditions of heavy-ion collisions (HICs) or is washed out by the violent dynamics. The answer carries important implications on the interplay between isospin dynamics and $np$ correlation: it bears on how neutrons and protons collectively transport in isospin-asymmetric nuclear matter and how $np$ correlations manifest in a hot, expanding, non-equilibrium environment. In ground-state nuclei, the evolution of the mean field and nucleon-nucleon correlations with mass, charge, and $N/Z$ asymmetry toward the drip lines is well studied \cite{Lee:2011zzb}. In HICs, by contrast, one can explore how these same properties, directly linked to nuclear forces, evolve with density and temperature as the neutron-to-proton asymmetry of the medium is modified. Such evolution with equation-of-state (EoS) variables is key to understanding supernova dynamics \cite{Oertel:2016bki} and neutron-star matter, including associated phenomena such as neutrino and gravitational-wave emission \cite{LIGOScientific:2017vwq, LIGOScientific:2018cki}. 

Focusing on tin isotopes, the ground-state charge radius of $^{132}$Sn differs from that of $^{108}$Sn by only about 3\% \cite{Angeli:2013epw, Gustafsson:2025xpd}, consistent with mean-field expectations. Theoretical studies have shown that $np$ correlations around the Fermi surface shape charge radius systematics across closed shells \cite{PhysRevC.109.064302}. If the collision dynamics simply reflected the ground-state distributions, one would expect a similarly modest difference in the proton emitting source size between neutron-rich and neutron-deficient systems. However, the dynamics of a HIC may  amplify correlations that are mild in the ground state. Transport models that incorporate the neutron skin and mean-field dynamics successfully describe many observables \cite{PhysRevC.110.064617}, yet they may not fully capture the dynamical amplification of $np$ correlations that can arise under the extreme compression and expansion of the collision zone. Thus, a direct comparison between neutron-rich and neutron-deficient Sn+Sn collisions offers a clean test of whether such beyond-mean-field effects emerge in the collision dynamics and whether they manifest in the spatial distribution of emitted protons.

Two experimental tools are essential to achieve this goal. First, collisions induced by radioactive heavy ions with widely different neutron excess provide a unique laboratory to probe the proton and neutron distributions under conditions of extreme isospin asymmetry and dynamic phase-space rearrangement. These conditions closely mimic the violent stellar environments for which the EoS of asymmetric nuclear matter must be reliably known \cite{Huth:2021bsp, Tsang:2023vhh, Sun:2022xjr, Sorensen:2023zkk}. Second, femtoscopy, historically known as intensity interferometry, offers an indispensable tool to image the space-time structure of particle-emitting sources \cite{Wang:2021mrv, PhysRevC113.014603, Xu:2024dnd, Zhang:2025yqq, Huang:2025edi} as well as to probe final-state interactions \cite{Si:2025eou, STAR:2015kha}. In particular, by measuring the proton-proton (p-p) correlation function (CF) in Sn+Sn collisions with large $N/Z$ differences, one can precisely extract the spatial distribution of the proton emitting source and infer how the system's neutron abundance influences the emitting region, shedding new light on the role of $np$ correlations in extremely dynamic collisions.

In this Letter, we report measurements of the p-p CF in two reaction systems with substantially different $N/Z$ ratios: $^{132}$Sn+$^{124}$Sn (neutron-rich) and $^{108}$Sn+$^{112}$Sn (neutron-deficient), at a beam energy of 270~MeV/nucleon, performed at the RIBF facility. Using a two-component source model, we extract the spatial distribution of the proton emitting source for these two systems with distinct isospin asymmetries. The observed difference in source size far exceeds the modest ground-state charge-radius difference, revealing a significant dynamical amplification. The implications of this finding for the role of $np$ correlations in the collision dynamics are discussed.

\begin{figure}[ht!]
  \includegraphics[width=0.90\linewidth]{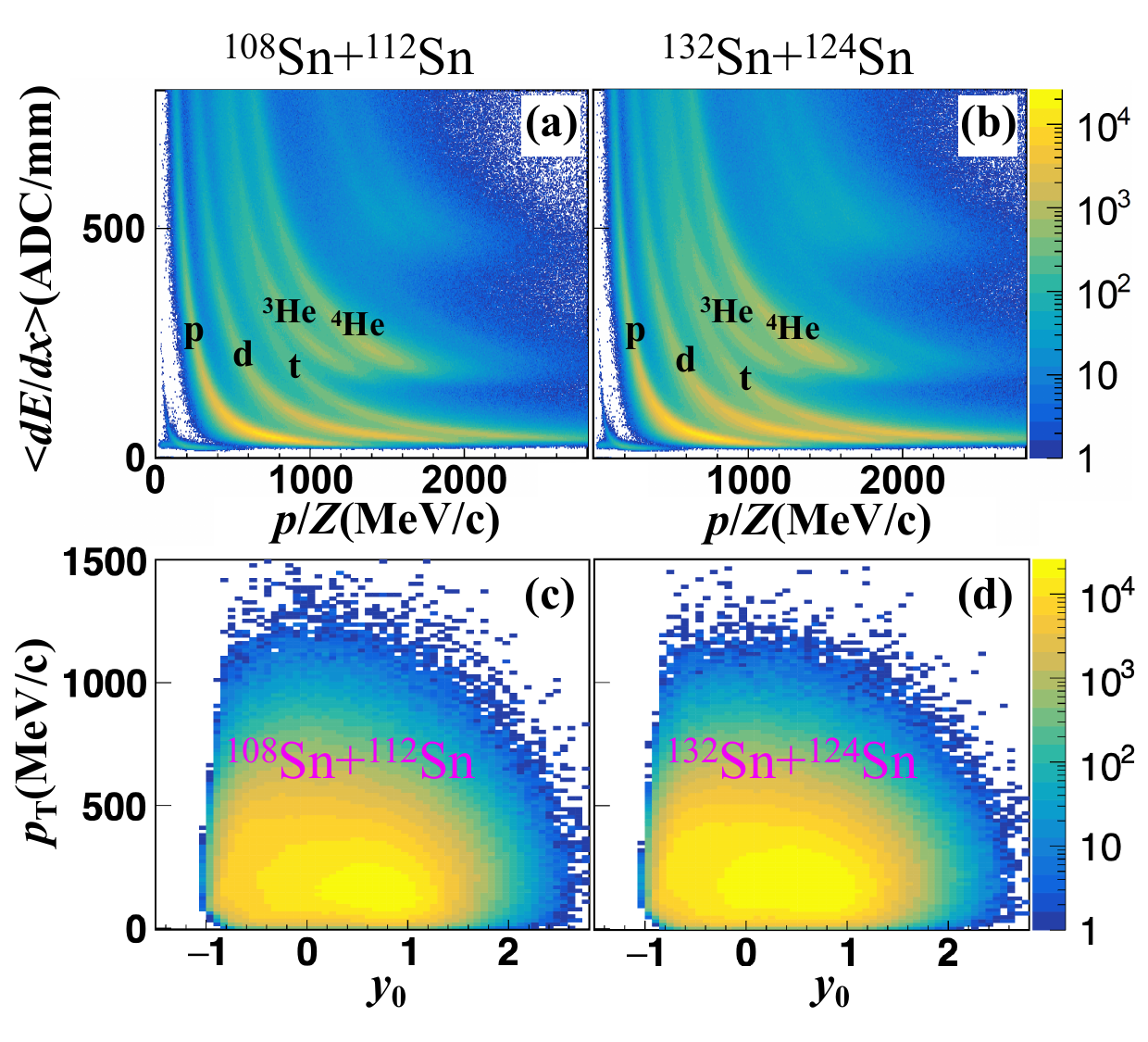}	
  \caption{Particle identification plot from $^{108}$Sn +$^{112}$Sn (a) and $^{132}$Sn +$^{124}$Sn (b) reactions, measured with the S$\pi$RIT TPC. Phase space distribution of protons detected in the experiment about scaled rapidity $y_{0}$ and transverse momentum $p_{\rm T}$ for $^{108}$Sn +$^{112}$Sn (c) and $^{132}$Sn +$^{124}$Sn (d) reactions.} 
  \label{fig:PID_Phase}
\end{figure}

The experiment was performed at the Radioactive Isotope Beam Factory at RIKEN, within the S$\pi$RIT collaboration \cite{SpiRIT:2021gtq, SpiRIT:2021och, SpiRIT:2022sqt, SpiRIT:2023htl, Kurata-Nishimura:2025ydt}. Beams of $^{108}$Sn and $^{132}$Sn projectiles at an incident energy of 270 AMeV bombarded the isotopically enriched ($>95\%$) $^{112}$Sn and $^{124}$Sn targets with areal densities of 561 and 608 mg/cm$^2$, respectively. Light charged particles were detected with the S$\pi$RIT time projection chamber (TPC) \cite{SpiRIT:2014yhq, SpiRIT:2016vpd, SpiRIT:2019jjg, Barney:2020mxk, Lasko:2016igj, Isobe:2018udc}, placed inside the SAMURAI spectrometer \cite{OTSU2016175}. Using the S$\pi$RITROOT software framework \cite{SpiRIT:2020gce, SpiRITGithub}, space-charge effects from the beam traversing the TPC were corrected \cite{SpRIT:2019xqz}. Hydrogen and helium isotopes were cleanly identified by their electronic stopping power $dE/dx$ and magnetic rigidity $B\rho = p/Z$ for both the $^{108}$Sn+$^{112}$Sn and $^{132}$Sn+$^{124}$Sn reactions, as shown in Fig.~\ref{fig:PID_Phase} (a) and (b), respectively.

We focus on central collisions. { These events are selected using the multiplicity of detected charged particles \cite {Wang:2026bog}. They correspond to an impact‑parameter ratio $b/b_{\rm max}$ $\le$ 0.5 (centrality: 0-25\%), where $b_{\rm max}=1.15\left(A_{\text{P}}^{1/3}+A_{\text{T}}^{1/3}\right)$. Here} $A_{\text{P}}$ and $A_{\text{T}}$ are the mass number of projectile and target, respectively \cite{SpiRIT:2021och,Barney:2020mxk,Myers:1973xlq}. The phase space distributions of protons in this analysis are shown in Fig.~\ref{fig:PID_Phase} (c) and (d) for $^{108}$Sn +$^{112}$Sn and $^{132}$Sn +$^{124}$Sn reactions, respectively. The horizontal axis is the  scaled rapidity $y_{0}=y/y^{\rm cm}_{\rm NN} - 1$, where $y$ is the laboratory rapidity of the detected particle and $y^{\rm cm}_{\rm NN}$ is the rapidity of the nucleon-nucleon (NN) center of mass in the laboratory frame.  For both reactions $y^{\rm cm}_{\rm NN} \approx 0.371$ \cite{SpiRIT:2021och}. The vertical axis is the transverse momentum $p_{\rm T}$. The phase-space distributions of protons are remarkably similar between the two reaction systems. 

The CF was constructed using the relative momentum $k^{*}=\left|\mathbf{p}_{1}-\mathbf{p}_{2}\right|/2$ of the two protons in their pair rest frame (PRF), where $\mathbf{p}_{1}$ and $\mathbf{p}_{2}$  denote the PRF momenta of the respective  particles. The experimental CF is defined as 
\begin {equation}
C_{\rm exp}(k^{*})=A\frac{N^{\rm same}(k^{*})}{N^{\rm mix}(k^{*})},
\label{eq1}
\end {equation}
where $N^{\rm same}(k^{*})$ is the relative momentum distribution of two particles originating from the same event, and $N^{\rm mix}(k^{*})$  is the reference distribution obtained via event mixing \cite{Lisa:1991zz,Kopylov:1974th}. The normalization parameter $A$ is determined by requiring  that $C_{\rm exp}(k^{*}) = 1$ for large relative momenta ($k^{*} > 80$ MeV/c). 

Identical single-particle cuts on track-quality  and mass gating for particle identification are applied
in { ``same'' and ``mixed''} events. The track splitting (one single particle reconstructed as two tracks) and track merging (two particles with similar momenta reconstructed as one track) effects are removed via the cuts on separate distance between two tracks \cite{STAR:2024zvj,STAR:2004qya}.

\begin{figure}[ht!]
  \includegraphics[width=0.8\linewidth]{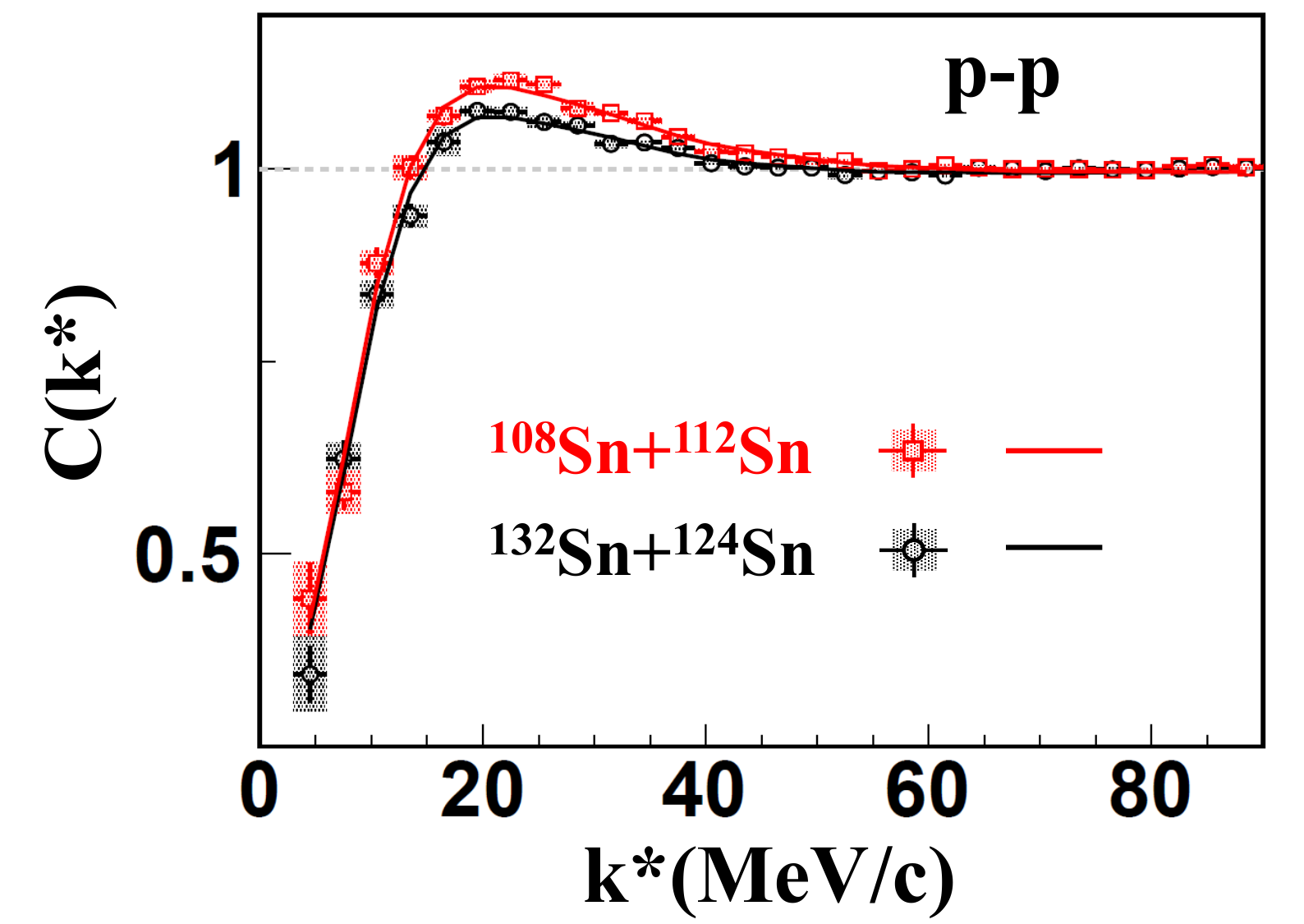}	
  \caption{Proton-proton correlation functions for the $^{108}$Sn +$^{112}$Sn (red open squares) and $^{132}$Sn +$^{124}$Sn (black open circles) reactions, respectively. The  vertical lines (shaded squares) indicate statistic (systematic) uncertainties. The solid curves correspond to the fitting results for $^{108}$Sn +$^{112}$Sn(red) and $^{132}$Sn +$^{124}$Sn(black), respectively. } 
  \label{fig:ppCF}
\end{figure}

{ The data points on Fig.~\ref{fig:ppCF} show} the experimental p-p CFs for the $^{108}$Sn +$^{112}$Sn and $^{132}$Sn +$^{124}$Sn reactions. Systematic uncertainties, represented by shaded squares, were evaluated by varying
the applied selection criteria, including track quality, mass gating, multiplicity, track splitting and track merging \cite{Wang:2026bog}. 
Comparing the CFs masured for the two reaction systems, we observe that the strength of the peak at $k^* \approx 20$ MeV/c is  {weaker} for $^{132}$Sn +$^{124}$Sn, as compared to $^{108}$Sn +$^{112}$Sn. The difference in the height of the peak amounts to about 0.04, far beyond the uncertainty. This result suggests that the neutron abundance of the colliding system significantly influences the p-p correlation strength. { This feature is also observed in central $^{40}$Ca+$^{40}$Ca  and $^{48}$Ca+$^{48}$Ca reactions at 80~MeV/nucleon\cite{Henzl:2011dh}.}
{ Moreover, the observed difference between the two systems remains robust against variations in rapidity, centrality, and direction selections\cite{Sup_Mat}.}

To extract the spatial information of the proton emitting source, the p-p CFs were analyzed using the angle-averaged Koonin-Pratt equation \cite{Nzabahimana:2023tab, Nzabahimana:2024knd, Tam:2025mkk}:

\begin {equation}
 C(k^{*})=4\pi\int K(k^{*},r)S(r)r^{2}dr, 
\label{eq2}
\end {equation}
where  $K(k^{*},r)$  is the kernel function that encodes the final-state interactions between the two particles,   and $S(r)$ is the emitting source function characterizing the relative distance $r$ between two particles. In the case of non-simultaneous emission of the pair, $r$ is calculated when the second particle is emitted. 

In our work, a two-component emitting source is adopted, consisting of a fast core component and a slow tail component. The fast core describes the small-$r$ region of the source, where proton pairs are predominantly at the early dynamical stage of the collision. The slow tail at large-$r$ values arises from pairs emitted by slowly evolving mechanisms, such as evaporation and secondary decay \cite{Nzabahimana:2024knd, Tam:2025mkk, Verde:2001md, Verde:2003cx}. The fast core and the slow tail are parameterized, respectively, as a Gaussian function and an exponentially decaying function \cite{Nzabahimana:2024knd}:


\begin {equation}
S(r)  = \dfrac{\lambda}{(2\sqrt{\pi}R_{c})^3} \exp\bigg(-\dfrac{r^2}{4R_{c}^2}\bigg) + \dfrac{1-\lambda}{96\pi T_{s}^5}r^2 \exp\bigg(-\dfrac{r}{T_{s}}\bigg).
\label{eq3}
\end {equation}

The \Rf parameter in the first Gaussian term is the radius (size) of the fast core source. $T_{s}$ characterizes the space-time extent of the exponential tail. By definition, the integral of the whole source function is equal to unity. Therefore, the $\lambda$ parameter is the integral of the fast core portion and represents the probability of emitting proton pairs at the early dynamical stage of the collision. 

\begin{table*}[ht]
\centering\caption{\label{tab:table1} The fitting results for the source parameters, including statistical and systematic errors.}
\renewcommand{\arraystretch}{1.1}
\renewcommand{\tabcolsep}{0.7pc}
\begin{tabular}{cccc}
\hline
  \hline
    Reactions & \Rf(fm) &  $\lambda$ & $T_{s}$(fm)    \\
  \hline
   $^{108}$Sn +$^{112}$Sn   & 1.74$\pm$0.08(stat.)$\pm$0.05(syst.)
   &   0.064$\pm$0.007(stat.)$\pm$0.004(syst.)
 &   7.53$\pm$0.21(stat.)$\pm$0.22(syst.)
     \\
   $^{132}$Sn +$^{124}$Sn   & 2.22$\pm$0.13(stat.)$\pm$0.07(syst.)
 &   0.076$\pm$0.011(stat.)$\pm$0.007(syst.)
 &   7.45$\pm$0.20(stat.)$\pm$0.20(syst.)
     \\
 \hline
  \hline
\end{tabular}
\end{table*}

We applied the well-developed imaging technique \cite{Nzabahimana:2024knd, Tam:2025mkk, Verde:2001md, Verde:2003cx, Brown:1997ku} to infer the spatial profile Eq. (\ref{eq3}) of the proton emitting source. 
 To fit the data, the kernel matrix $K(k^{*},r)$ in Eq. (\ref{eq2})
 was computed by solving the Schrödinger equation with known p-p interaction \cite{Tam:2025mkk}. The loss function is defined as the $\chi^{2}$ value between the fitted CF curve and the experimental data over the relative momentum range $k^* \in [3,90]$ MeV/c. Using a gradient-descent optimization, the best-fit CF curves are obtained and plotted on top of the data points, as shown in Fig.~\ref{fig:ppCF}. The corresponding source parameters, along with their statistical and systematic uncertainties, are listed in Table~\ref{tab:table1}. 

\begin{figure}[ht!]
  \includegraphics[width=0.67\linewidth]{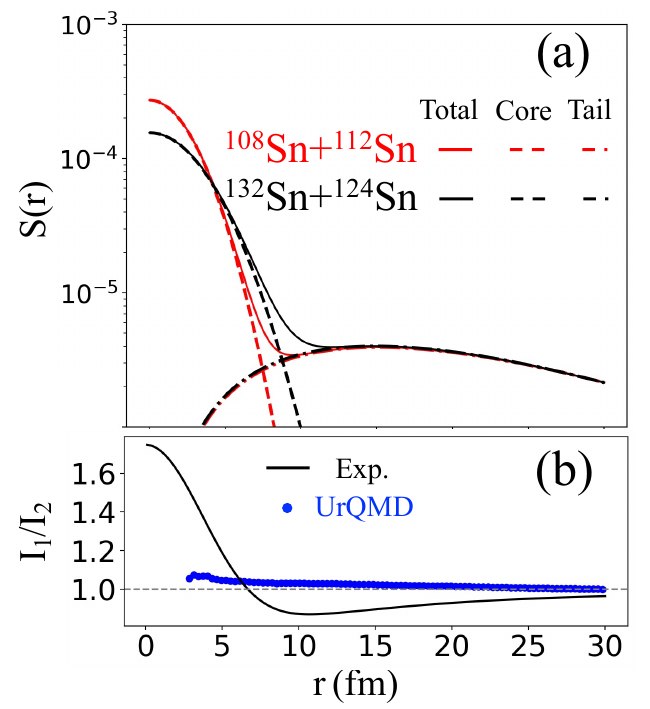}	
  \caption{(a) Fitting results of source function for the $^{108}$Sn +$^{112}$Sn (red) and $^{132}$Sn +$^{124}$Sn (black) reactions. Solid lines: total distribution; dashed lines: core component; dash-dotted lines: tail distribution. (b) Ratio of the integral of the source distribution for $^{108}$Sn +$^{112}$Sn to that for $^{132}$Sn +$^{124}$Sn. { The fit to the experimental data is shown by the black solid line, while the UrQMD results are represented by blue dots.}} 
  \label{fig:fit_result_Sr}
\end{figure}

 
The profile of the source function, $S(r)$, extracted from fitting the data for the two collision systems, are shown as { solid} lines on Fig.~\ref{fig:fit_result_Sr} (a). We also show, separately, the fast and the slow components as dashed and a dot-dashed lines, respectively. 
In the neutron-deficient system ($^{108}$Sn+$^{112}$Sn), the fast component is notably more compact, as compared to the case of the neutron-rich system ($^{132}$Sn+$^{124}$Sn): the size parameter $R_c$ is indeed observed to be smaller in the neutron-deficient collision (see Table~\ref{tab:table1}). In contrast, in the profile of the slow emission tail we do not show significant differences bewteeen the studied reaction systems, suggesting that the long-range component of the emitting source may be governed by similar mechanisms irrespective of the initial isospin asymmetry. Such extended tails have also been observed in relativistic HICs through deblurring imaging \cite{Xu:2024dnd}, L\'evy source fitting \cite{Kincses:2024lnv, Tsallis:1995zz, PHENIX:2024vjp}, and machine-learning approaches \cite{Wang:2024bpl}.

Besides the source sizes, a key feature of the source profile is its integral over the small-r region. This integral is predominantly governed by the correlation function strength near the 20 MeV/c peak, which directly characterizes the proton-pair emission probability at the early collision stage. For small-r values ( $r <$ 5 $fm$), the proton-rich system exhibits a distinctly enhanced source strength.  We compute the cumulative integral of the source function as $I(r)=4\pi \int_{0}^{r} r'^2 S(r') dr'$ for both reaction systems. Figure~\ref{fig:fit_result_Sr}(b) presents the integral ratio $I_{1}/I_{2}$, where $I_1$ and $I_2$ correspond to the proton-rich and proton-deficient systems, respectively, as a function of the integration upper limit $r$. The ratio $I_{1}/I_{2}$ starts at approximately 1.7 for $r\approx$ 0 $fm$ and declines to 1.2 at $r\approx$ 5 $fm$, demonstrating considerably stronger early-stage proton-pair emission in the proton-rich system. The ratio subsequently drops below unity with an inversion near $r\approx$ 6.5 $fm$, and asymptotically converges back to unity at large $r$. This behavior indicates negligible isospin dependence in the line shape of the slow source profile. These findings provide critical insights into the isospin-modulated proton-pair emission mechanism, which requires further validation via transport model simulations, an investigation beyond the scope of the present work. Instead, we explore this isospin effect by analyzing the differences in fast source sizes between the two collision systems.




\begin{figure}[ht!]
  \includegraphics[width=0.67\linewidth]{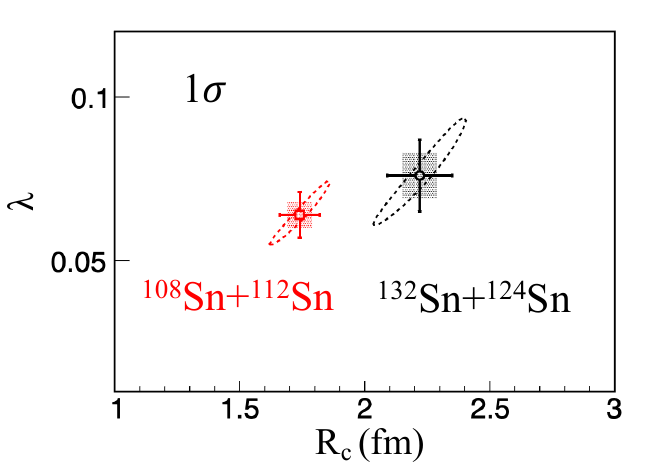}	
  \caption{ Fitting results for the $^{108}$Sn +$^{112}$Sn (red open squares) and $^{132}$Sn +$^{124}$Sn (black open circles) reactions, with the solid error bars (shaded squares)  indicating the statistical (systematic) uncertainties.  The dashed curves correspond to 1 $\sigma$ contour of the fitting.} 
  \label{fig:fit_result}
\end{figure}


The quantitative comparison of the fit parameters (whose numerical values are reported on Table~\ref{tab:table1}) is summarized in Fig.~\ref{fig:fit_result}. The best-fit values of the fast core radius $R_c$ and the fraction parameter $\lambda$ are shown with their statistical (error bars) and systematic (shaded squares) uncertainties. The $1\sigma$ confidence contours are indicated by the dashed curves. The fraction $\lambda$ remains around 7\% for both reactions, with a slightly larger value observed for the neutron-rich system. The contours reveal a positive correlation between $R_c$ and $\lambda$. Nevertheless, even when this correlation is taken into account, the core radius of the neutron-rich system is significantly larger than that of the neutron-deficient system. 


The extracted fast emission core radii are $R_c = 1.74 \pm 0.08\ \text{(stat.)} \pm 0.05\ \text{(syst.)}$~fm for the $^{108}$Sn+$^{112}$Sn reaction and $2.22 \pm 0.13\ \text{(stat.)} \pm 0.07\ \text{(syst.)}$~fm for the $^{132}$Sn+$^{124}$Sn reaction (Table~\ref{tab:table1}). Combining the statistical and systematic uncertainties in quadrature, the two values differ at a confidence level of 99.4\%. The relative difference in $R_c$ between the two reactions is approximately 24\% (relative to the average value). This is remarkably large in comparison to the ground-state properties in the entrance channel of the studied reactions. Indeed, the $\langle r_{\text{ch}}^2 \rangle$ of the involved tin isotopes have been determined as 4.5605~fm for $^{108}$Sn, 4.5948~fm for $^{112}$Sn, 4.6735~fm for $^{124}$Sn, and 4.7093~fm for $^{132}$Sn \cite{Angeli:2013epw,Gustafsson:2025xpd}. The  difference of $\langle r_{\text{ch}}^2 \rangle$ between the two projectiles $^{132}$Sn and $^{108}$Sn is 3.2\%.  When taking the average of projectile and target, it reduces to approximately 1.2\% between the two reactions. Strikingly, the observed 24\% difference in $R_c$ exceeds the ground-state charge‑radius difference by an order of magnitude.


A direct explanation for this amplified difference might invoke the pulling effect of excess neutrons located in the nuclear periphery on protons, driven by isospin diffusion \cite{Tsang:2008fd, Chen:2004si} (a mean-field effect). This effect can be estimated from the mean transverse momentum. Experimentally, $\langle p_{\text{T}}\rangle = 251.7$~MeV/$c$ is obtained for the $^{108}$Sn+$^{112}$Sn reaction and $256.6$~MeV/$c$ for the $^{132}$Sn+$^{124}$Sn reaction. Such mean-momentum (velocity) difference causes only 2\% difference to the proton emitting source size, and it is much smaller than the $R_c$ difference observed from our study of p-p CFs.  
A critical closure test is provided by simulations using UrQMD transport model \cite{Li:2025mox}: with {conventional mean-field dynamics  frame}, the expected difference in the proton emitting source size between the two systems is only at the few-percent level, {  as shown by the ratio $I_1/I_2$ in Fig. \ref{fig:fit_result_Sr} (b), slightly} depending on the symmetry energy parameterization. 

The fact that such simulations based merely on a mean-field scheme cannot reproduce the observed 24\% difference strongly indicates that the amplification points to beyond-mean-field origins. An earlier theoretical work by Wei \textit{et al.}~\cite{PhysRevC.101.014613} studied the p-p CF in $^{197}$Au+$^{197}$Au collisions at 400~MeV/nucleon and found that incorporating short-range correlations (SRCs) of 12\% abundance in the initial nucleon momentum distribution led to a decrease of p-p correlation strength by approximately 0.1.  
{Here, SRC refers to correlated nucleon pairs generated by the attractive tensor‑force component of the nucleon‑nucleon interaction. Such correlation manifests itself in the high‑momentum tail of the nucleon‑momentum distribution and predominantly involve neutron‑proton pairs. 
}
Accordingly, from the 0.04 difference of the p-p CF strength in Fig. \ref{fig:ppCF}, one naively expects  4.8\% more protons forming $np$ SRC pairs in $^{132}$Sn+$^{124}$Sn than in $^{108}$Sn+$^{112}$Sn. These results consistently demonstrate that $np$ correlations enlarge the proton emitting source size extracted from  p-p CFs. Later, in an independent transport simulation of two-proton knock-out reaction of $^{12}{\rm C}(p,2p)^{11}{\rm B}$, the short-range repulsion (to mimic SRC) similarly weakens the p-p CF by extending the system size \cite{PhysRevC.105.014603}.   

The observation of the amplified difference in $R_c$ between the two reactions provides an intriguing signature that short-range correlations can influence long-range nuclear properties not only in the ground state (such as charge radii), as proposed in Ref.~\cite{MILLER2019360}, but also in the dynamic environment of  HICs.  Complementary to our observation, the hard bremsstrahlung $\gamma$ spectrum from $^{124}$Sn+$^{124}$Sn reactions has been measured, clearly demonstrating that the signal of $np$ SRCs in the initial ground state of $^{124}$Sn survives the collision process \cite{Xu:2025mvv, Xu:2026prc}. 
Notably, since the emitted protons experience the compression stage, the two-body correlations are not only encoded in the initial state, but also play a role in the collision  process. To better describe final observables in HICs, it is necessary to incorporate two-body correlations explicitly in modern transport models. Leveraging the precision of the femtoscopic techniques, our result serves as a critical benchmark for the development of such models, moving beyond the single-particle (mean-field) picture.

In summary, we have measured p-p CFs in collisions induced by radioactive ion beams, comparing two reaction systems with markedly different neutron-to-proton ratios, $^{132}$Sn+$^{124}$Sn (neutron-rich) and $^{108}$Sn+$^{112}$Sn (neutron-deficient), at a beam energy of 270~MeV/nucleon. Leveraging the sub-femtometer spatial resolution inherent to the correlation function method, it is found that the radius parameter of the dynamic proton emitting source function is remarkably larger, by about 24\%, in the neutron-rich system than in the neutron-deficient system. This difference is about an order of magnitude larger than the difference that one would expect by studying ground state charge radii. Such observation cannot be explained by transport model simulations based on a mean-field picture, indicating that the observed $N/Z$ dependence of $R_c$ originates from beyond-mean-field effects, specifically short-range $np$ correlations.

Our findings support the conclusion that  short-range correlations permeate a wide range of observables, including those that are accessible in relativistic HICs through fine probes and precise analysis of femtoscopy with proton-proton correlation functions. { Notably, similar effect of SRC on p-p CFs is reported in a very recent simulation of O+O at top RHIC energy \cite{Xi:2026vrp}.} Since HICs  represent the only terrestrial means to produce nuclear matter away from saturation density, we believe that the observed effect of $np$ correlations on the proton emitting source radii carries important information that need to be taken into account. Such correlations must be explicitly included as a beyond-mean-field effect in our efforts to constrains the EoS of asymmetric nuclear matter with both HICs and neutron-star observations \cite{Li2026epj, Li:2026qvl, Sedrakian:2024uma, Cai2026, Nzabahimana:2025ivc}.

{\it Acknowledgements -} The authors  thank Prof. Bao-An Li and Prof. Gaofeng Wei for fruitful discussions, and Dr. Pengcheng Li for providing UrQMD simulation data. 
Computing resources were provided by the HOKUSAI-Great Wave system at RIKEN, and the Center for High Performance Computing in Tsinghua University. 
This work was supported by the Nation Science Foundation of China under grant No.12205160 and 12335008, the Ministry of Science and Technology of China under Grant No. 2022YFE0103400, the Japanese MEXT KAKENHI
(Grant-in-Aid for Scientific Research on Innovative
Areas) grant No. 24105004, JSPS KAKENHI Nos.
JP17K05432 and JP19K14709, and JP21K03528, the U.S.
Department of Energy under Grant Nos. DE-SC0014530,
DE-NA0002923, and DE-FG02-93ER40773, the US National
Science Foundation Grant No. PHY-1565546, the
Polish National Science Center (NCN), Poland, under contract
Nos. UMO-2013/09/B/ST2/ 04064 and
UMO-2013/10/M/ST2/00624, and the National Research
Foundation of Korea (NRF) under Grant Nos.
2018R1A5A1025563, RS-2024-00333673 and RS-2024-00436392.
I. G. was supported by HIC for FAIR and the
Croatian Science Foundation under projects Nos. 1257
and 7194.
Z. G. Xiao acknowledges the support from the Initiative Scientific Research Program of Tsinghua University. 

%

\bibliography{references.bib}

\end{document}